\documentclass[conference]{IEEEtran}
\usepackage{amsmath}
\usepackage{amsthm}
\usepackage{amssymb}
\usepackage{booktabs}
\usepackage{ifthen}
\usepackage{url}
\usepackage{listings, lstautogobble}
\usepackage{xspace}
\usepackage{graphicx}
\usepackage[most]{tcolorbox}
\usepackage{xcolor}

\definecolor{myhighlight}{RGB}{255,255,102} 

\usepackage{wrapfig}

\usepackage{hyperref}
\usepackage[nameinlink]{cleveref}

\newcommand{\projName}{\textsc{Tecton}\xspace} 
\newcommand{\bosque}{\textsc{Bosque}\xspace} 

\newcommand{\eg}{\hbox{\emph{e.g.}}\xspace}
\newcommand{\ie}{\hbox{\emph{i.e.}}\xspace}
\newcommand{\etc}{\hbox{\emph{etc.}}\xspace}
\newcommand{\vs}{\hbox{\emph{vs.}}\xspace}

\newcommand{\etal}{\hbox{\emph{et al.}}\xspace}

\newcommand{\aka}{\hbox{\emph{a.k.a.}}\xspace}

\newcommand{\cf}[1]{\texttt{#1}} 

\newboolean{showcomments}
\setboolean{showcomments}{true} 
\ifthenelse{\boolean{showcomments}}{

}{
}

\newboolean{addlater}
\setboolean{addlater}{false}
\ifthenelse{\boolean{addlater}}{
  \newcommand{\addlater}[1]{#1}
}{
  \newcommand{\addlater}[1]{}
}

\definecolor{cgreen}{rgb}{0.25,0.5,0.35} 
\definecolor{stringred}{rgb}{0.6,0,0} 

\lstdefinelanguage{bosque}{
keywords={function, api, type, of, requires, ensures, softchk, var, let, return, if, elif, else, assert, invariant, field, pred, fn, entity, env, resources, datatype},
keywordstyle=\color{blue}\bfseries,
identifierstyle=\color{black},
alsoother={@},
sensitive=true,
comment=[l]{\%\%},
commentstyle=\color{cgreen}\bfseries,
string=[b]{"},
stringstyle=\color{stringred}\ttfamily,
}
 
\lstdefinelanguage{typespec}{
keywords={scalar, model, extends, union, boolean, int32, string, op, @discriminated, @minValue, @maxValue, @pattern},
keywordstyle=\color{blue}\bfseries,
identifierstyle=\color{black},
sensitive=true,
comment=[l]{\/\/},
commentstyle=\color{cgreen}\bfseries,
string=[b]{"},
stringstyle=\color{stringred}\ttfamily,
}
 
\usepackage{algorithm}
\usepackage{algpseudocode}
\newcommand{\myalgofonts}{\footnotesize}

\begin{document}

\title{Independent Test Generation for RESTful APIs}

\author{
\IEEEauthorblockN{
S M Sadrul Islam Asif\IEEEauthorrefmark{1},
James Chen\IEEEauthorrefmark{2},
Kennett Puerto Diaz\IEEEauthorrefmark{1},
Earl T. Barr\IEEEauthorrefmark{3},
Mark Marron\IEEEauthorrefmark{1}
}
\IEEEauthorblockA{\IEEEauthorrefmark{1}University of Kentucky, Lexington, Kentucky, USA\\
Email: \{sadrul.islam, Kennett.Puerto, mark.marron\}@uky.edu}
\IEEEauthorblockA{\IEEEauthorrefmark{2}Meta, Menlo Park, California, USA\\
Email: jameszc@meta.com}
\IEEEauthorblockA{\IEEEauthorrefmark{3}University College London, London, UK\\
Email: e.barr@ucl.ac.uk}
}

\maketitle
\begin{center}
\small
\textit{This paper was previously posted on arXiv under the title
``BOSQTGEN: Breaking the Sound Barrier in Test Generation''.}
\end{center}

\vspace{0.5em}
\begin{abstract}
	Modern REST API testing relies on brittle sequences of calls to build system state. These multi-step tests suffer from non-determinism, poor scalability, and a ``reachability tax'' where a single failed setup step invalidates the entire test. We introduce \projName, which breaks this cycle by replacing implicit state construction with explicit state synthesis of both the request payload and mock data it depends on.

\projName achieves this through two complementary mechanisms: it generates diverse, valid payloads directly, and it augments existing test mocks with realistic data so those payloads have valid system state to reference. Both mechanisms apply combinatorial testing to a new domain: the nested property space of Abstract Data Types (ADTs). \projName decomposes complex API requests into primitive components to unleash LLMs on the more tractable subtasks of identifying equivalence classes of these primitives and generating representative values for them. It then uses LLMs to extract and inject state values via test mocks, enabling payloads to reference valid state. It recomposes these values into covering combinations to directly produce high-coverage test payloads.

On standard RESTful benchmarks, \projName achieves 70\% average line coverage --- a 20\% absolute increase over sequence-based generators.  It exposes 2× more runtime errors than any prior tool, including assertions and data constraint failures. \projName's shift from sequencing API calls to synthesized payloads advances the state of the art in automated API validation.

\end{abstract}


\section{Introduction}
\label{sec:intro}

Software engineering is undergoing a structural
transformation, moving from monolithic codebases to modular, API-centric
architectures. In some cases,
APIs serve simply as a mechanism for managing complexity by decomposing the
problem; in many cases, however, they represent a new economic model of
development focused on composition and integration of existing code or
services. The appeal of adopting API centric development is clear:  it facilitates reuse and the
reorganization of developers into small and agile teams, aligned around single
responsibilities. These teams, combined with extensive, API-unlocked re-use,
enable developers to focus exclusively on the core functionality of their
features or components~\cite{twopizza,conway}.  

For these reasons, more and more applications now decompose functionality into
a set of services that communicate via well-defined APIs and often use
RESTful~\cite{restphd} architectures. This shift elevates the API (RESTful service) 
contract to a critical role. Inadequate contracts --- whether due to weak type systems, 
ambiguous documentation, or insufficient testing --- lead to mismatched expectations. 
These gaps can produce outcomes ranging from simple crashing bugs
(e.g., confusion over optional fields) to severe security failures or data
breaches. Closing these gaps via conformance testing motivates the need for
robust API test generation.

While the original REST formulation~\cite{restphd} is built on a stateless
ideal, the reality of modern development often defaults to a ``pseudo-stateful''
model where the application logic assumes a specific lifecycle: endpoints can
only be successfully invoked after a chain of preceding calls has
populated the server-side state. To tackle this, prior state-of-the-art tools
like RESTler~\cite{restler} rely on fuzzing sequences of calls to dynamically
discover which state transitions are valid.

We argue that this reliance on sequences imposes a ``reachability
tax’’ that makes testing inherently brittle and expensive. Because each test
case depends on a hidden and implicit chain of previous calls, the suite
inherits the cumulative probability of failure from every setup step.
Non-determinism or changes in test order can lead to entirely different
outcomes, a problem exacerbated by the cost of long setup sequences required to
reach deep logic.

We introduce \projName, a methodology that achieves \emph{state-aware
statelessness}. Rather than treating server-side state as an implicit
side-effect of a sequence, \projName treats it as a first-class, synthesizable
artifact. By pivoting from implicit state construction to explicit, single-call
state synthesis, \projName restores the stateless ideal of REST while maintaining
the semantic depth required to test complex logic. To do so effectively, \projName
must build valid and diverse payloads and provide the tested API endpoint with
a valid execution environment in which to handle that payload.  The problem of
constructing and sampling instances of abstract data types (ADTs), \aka the
structured input generation problem, has hindered payload construction.
To solve it, \projName
1) decomposes complex, dynamic ADT parameters into primitive types,
2) partitions these primitives and selects representative values, and
3) efficiently selects and combines these values into diverse instances.

To decompose ADTs into their constituent primitive types, \projName recursively
traverses them (\Cref{sec:decompose}), then, rather than relying on domain experts
or random sampling, it uses an LLM to partition them and generate
representative values for them (\eg, a Fahrenheit temperature)
(\Cref{sec:llm-partitions}). Alone, decomposition and partitioning are not
enough: even a modest set of values explodes combinatorially. To solve this
problem, \projName leverages the sparsity-of-effects principle from
statistics~\cite{wu2021experiments} that underlies combinatorial testing (CT).
While CT successfully applies this principle to program parameters,
\projName is its first application to the nested property-space of complex Abstract
Data Types (ADTs). 

To allow its its ``stateless'' payloads interact with a valid and consistent
environment, \projName uses data from existing human-authored test mock 
datasources (\eg databases) if they exist or it can auto-populate a mock datasource 
with LLM-generated values. \projName then extracts and injects these values directly 
into LLM prompts. This novel approach to handling external state allows generated 
payloads to reference valid existing entities (\eg, a
user ID from a mock database), enabling state-dependent code paths without
multi-step setup sequences (\Cref{sec:mockgen}). 

Unlike prior work leveraging LLMs to tackle API testing, \projName restricts LLMs
to narrow, well-defined tasks~\cite{kang-etal-2024-empirical}: decomposing
primitive types into equivalence classes and generating representative values,
and relevant mock data. Each LLM task is carefully scoped to a low-dimensional,
distributionally narrow problem, keeping the model's role tractable (inference cost is under $\$0.10$ per application) and the
overall approach robust to model changes or failures. This disciplined
decomposition of responsibility --- rather than asking a single LLM to
synthesize entire test payloads or sequences in a single step --- is a key
methodological contribution of \projName.

We evaluate \projName head-to-head against the current state-of-the-art on the
canonical suite of RESTful
benchmarks from~\cite{arat-rl,autoresttest}: the de facto evaluation standard for
this domain. \projName achieves a commanding average line coverage of 70\% ---
representing a 20\% increase over existing sequence-based generators --- while
exposing 2$\times$ more runtime errors. These results suggest that \projName's state-aware statelessness offers a
robust alternative for uncovering deep logic failures that may be obscured or
rendered unreachable by the brittleness of traditional sequence-based
exploration.

\subsection*{Contributions}

\begin{itemize}

   \item We introduce \projName, a novel, \emph{state-aware stateless} test
       generation methodology that replaces brittle call-sequences with
       explicit state-synthesis. We show that by combining partitioned argument
       construction with contextual state-injection, we achieve a 2$\times$
       increase in error detection over current sequence-based generators
       (\Cref{sec:topline}).

   \item We are the first to apply the sparsity-of-effects principle to the
       internal property-space of ADTs. We demonstrate that by isolating and
       sampling the interactions of primitive constituents within nested
       structures, we can achieve high coverage without the reachability tax
       of multi-step sequences.

   \item We specify and realise a novel refinement of the LLM input generation
       problem by tasking the LLM with generating equivalence classes for
       primitive constituents rather than attempting to synthesize entire
       complex objects.

\end{itemize}


\section{\projName API Testing}
\label{sec:algorithm-overview}

We first describe how \projName leverages TypeSpec as its API specification language (\Cref{sec:typespec}), then overview the complete \projName algorithm for test generation, presenting how the core algorithm that decomposes complex types into primitives, generates representative values, and composes them via $k$-way sampling (\Cref{sec:core:alg}). We then show results that it produces on \Cref{fig:api-spec-example}. Additional details on compositional argument construction, LLM-based value generation, and mock data integration appear in later sections.

\subsection{\projName Leverages TypeSpec}
\label{sec:typespec}

The \projName tool operates on API specifications written in TypeSpec~\cite{typespec}, which was developed and is deployed in Azure services at Microsoft as a replacement 
for OpenAPI\footnote{Amazon AWS supports TypeSpec for RESTful service development via their Project Development Kit~\cite{awspdk} as well.}. TypeSpec 
provides a rich type system for defining structured data definitions while also providing seamless inter-operation with OpenAPI based systems. Given these characteristics, and 
industrial uptake, TypeSpec provides an ideal basis for exploring test input generation for RESTful APIs.

The running-example service specification in \Cref{fig:api-spec-example} shows a TypeSpec weather forecast API definition. This API takes a weather 
forecast as input and returns a list of recommended activities based on the conditions. Like many REST-style APIs, it relies on strings and numbers 
to represent a range of concepts, such as temperature and percentages. In TypeSpec additional constraints 
can be added as annotations or metadata on structure declarations as well. 

\begin{figure}[t]
\centering
\begin{lstlisting}[language=typespec,
numbers=left,
numberstyle=\tiny,
xleftmargin=0.04\linewidth]
scalar Fahrenheit extends int32;
scalar Mph extends int32;

@minValue(0) @maxValue(100)
scalar Percentage extends int32;

model TempRange { low: Fahrenheit; high: Fahrenheit; }
model WindSpeedRange { min: Mph; max: Mph; }

model Sunny { }
model Cloudy { }
model Precip { stormWatch: boolean; }

@discriminated
union ForecastInfo {
    sunny: Sunny;
    cloudy: Cloudy;
    precip: Precip;
}

model Forecast {
    temp: TempRange;
    windSpeed: WindSpeedRange;
    info: ForecastInfo;
    hourly: Array<Percentage>;
}

op recommendedActivities(v: Forecast): Array<string> {
    // Implementation details...
}
\end{lstlisting}
\caption{Weather forecast API specification in TypeSpec~\cite{typespec}.}
\label{fig:api-spec-example}
\vspace{-3mm}
\end{figure}

In \Cref{fig:api-spec-example}, several primitive types are \emph{type-declared} into distinct types, using the \cf{scalar} keyword, with additional constraints provided as 
annotations. The \cf{Fahrenheit} type is simply given a new name, but retains the dynamic range and operations of the underlying \cf{int32} type. The percentage type is 
restricted to values from $0$ to its max value of $100$ using the \cf{@minValue} and \cf{@maxValue} annotations. The \cf{TempRange} and \cf{WindSpeedRange} types are models (records) 
with properties that define structured data values. The \cf{ForecastInfo} datatype is a sum type, here representing different weather conditions, such 
as sunny, cloudy, or precipitation with a storm watch. Finally, the \cf{Forecast} model combines these types into a comprehensive structure that includes temperature, wind 
speed, and weather information, along with a list of hourly precipitation percentages. 

This structured approach allows for a clear and type-safe representation of the weather forecast API, enabling validation and testing of the API's behavior. 
OpenAPI specifications, as well as JSON Schema and Java/Python/JavaScript SDKs, can be automatically generated from TypeSpec models.

\subsection{\projName: The Core Algorithm}
\label{sec:core:alg}

A key challenge in applying sub-structural testing to real-world APIs is the prevalence of unbounded collections, like arrays, or inductive data-structures, such as trees. Partititioning simple types, like integers, or direct wrappers, like the Fahrenheit type in \Cref{fig:api-spec-example}, is straightforward. An integer might be partitioned into positive, negative, and zero values, yielding represenativies like $\lbrace 5, 0, -3 \rbrace$. A Fahrenheit type might be partitioned into sub-freezing, freezing, cool, warm, and hot ranges yielding $\lbrace -10, 32, 60, 80, 115 \rbrace$.  

Now consider the running example from \Cref{fig:api-spec-example}, which takes a weather forecast and returns an activity recommendation. This input includes both composite structures and collections, such as the array of hourly precipitation percentages.  For the \cf{hourly} array where partitioning the input space into a finite set is not obvious --- due both to the complexity of the data and the potentially large (unbounded) number of parts and values that could be selected. Should we include monotonically increasing arrays? Arrays with random variations? All identical values? Boundary cases like an empty array or all zeros? It becomes difficult, if not impossible, to enumerate all relevant parts and values with confidence.

Humans are quickly overwhelmed by the size and complexity of this input partitioning problem. AI agents (LLMs) have shown substantial successes in test 
generation~\cite{llmtestoverview,aitestingmeta,fuzz4all}, so one might hope that they could help. Yet they, too, struggle to reason 
about and generate a comprehensive covering set and representative values for complex structures. These problems are further compounded by the fact that LLMs have a 
limited context window, which with larger values, increases the likelihood of coherence loss and hallucination.

The solution has three steps: 1) decompose a complex ADT into its primitive components, 2) generate representative values for each primitive; and 3) combine those values via $k$-way sampling.  
\Cref{alg:gen-tests} realises this strategy to generate tests for API endpoints.

\Cref{alg:gen-tests} 
takes the type, $t$, of some parameter\footnote{
For simplicity, we focus on generating tests for one parameter. The \projName 
implementation extends to multiple parameters by grouping k-parameters into a composite k-component record type.}. \projName extracts all primitive
components from the type $t$ using the \textsc{getComponents} function
(\Cref{alg:getcomponents} in \Cref{sec:decompose}). For each identified primitive component, \projName
invokes \textsc{GenerateValues} to partition the primitive and obtain a set of candidate representative values.
These values may be generated in a variety of ways, including random sampling,
expert selection, constants mined from the application source, or, as we
show in \Cref{sec:llm-partitions}, using large-language models. \projName's core algorithm
then calls \textsc{GenKWay} (\Cref{alg:gen-comb-test} in \Cref{sec:soep}) to 
$k$-way sample these ADT primitive components and generate testing values.

\begin{algorithm}[t]
\caption{Generating full set of tests for a single API (one argument) with \projName.}
\label{alg:gen-tests}
\begin{algorithmic}
    \myalgofonts
\Function{testGen}{\emph{t}: Type} 
    \State components $\gets$ \Call{getComponents}{t, 'v'}
    \ForAll{c $\in$ components}
        \State c.values $\gets$ \Call{generateValues}{c} \Comment{Value generation$\dagger$}
    \EndFor

    \State tests $\gets$ $\emptyset$
    \ForAll{c$_1$...c$_k$ $\in$ components} \Comment{$k$-way Component Selection}
	\State tests $\gets$ tests $\cup$ \Call{GenKWay}{c$_1$, $\ldots$, c$_k$, components}
    \EndFor
    \State \Return tests
\EndFunction
\end{algorithmic}
\end{algorithm}

We now use \Cref{alg:gen-tests} to generate test values for the \cf{TempRange}
model type in \Cref{fig:api-spec-example}. It first decomposes the type into
the set of primitive components. The \cf{TempRange} model has two fields,
\cf{low} and \cf{high}, both of which are of type \cf{Fahrenheit}. Thus,
\textsc{getComponents} recursively decompose the model (\aka record) into two
primitive types, \cf{low} and \cf{high}, and then again recursively identify
these input components as primitive \cf{int32} types. The synthetic names for
these components are \cf{v.low.value} and \cf{v.high.value}. \projName then calls
\textsc{GenerateValues}, which uses an LLM to generate diverse,
domain-relevant, and finite set of values: $\lbrace -10, 0, 42 \rbrace$ for
\cf{v.low.value}  and $\lbrace 32, 60, 80 \rbrace$ for \cf{v.high.value}.

\begin{table}[t]
\caption{Example test cases for the \cf{TempRange} model (record) by 2-way sampling
$\lbrace -10, 0, 42 \rbrace$ and 
$\lbrace 32, 60, 80 \rbrace$.
}
\label{tab:temp-range-test-cases}
\centering
\ttfamily
\footnotesize
\begin{tabular}{ccc}
\toprule
\textbf{v.low.value} & \textbf{v.high.value} & \textbf{TempRange} \\
\midrule
-10 & 32 & $\lbrace -10, 32 \rbrace$ \\
-10 & 60 & $\lbrace -10, 60 \rbrace$ \\
-10 & 80 & $\lbrace -10, 80 \rbrace$ \\
\ldots & \ldots & \ldots \\
42 & 32 & $\lbrace 42, 32 \rbrace$ \\
42 & 60 & $\lbrace 42, 60 \rbrace$ \\
42 & 80 & $\lbrace 42, 80 \rbrace$ \\
\bottomrule
\end{tabular}
\end{table}

Using these values and $2$-way sampling, \projName finally calls \textsc{GenKWay} to create the test cases for the composite \cf{TempRange} model (\aka record) as 
shown in \Cref{tab:temp-range-test-cases}. This sampling generates a complete set of test 
cases that cover all combinations of the finite representative cases produced by the initial stages of \Cref{alg:gen-tests}. 
As we discuss in \Cref{sec:soep}, the sparsity of effects principle justifies why testing all k-way combinations (here $k=2$) of these primitive representatives suffices to expose most faults.

\section{\projName Decomposes Dynamic ADTs}
\label{sec:decompose}

Decomposition precedes partitioning. Before we can ask an LLM to partition
primitive types and generate representative values (\Cref{sec:llm-partitions}),
we must first extract those primitives from complex ADTs. \projName's
\textsc{GetComponents} (\Cref{alg:getcomponents}) recursively traverses a
type structure to identify all primitive components:  properties, array
entries, or scalar declarations of primitive types (numbers, strings, booleans, etc.).
These primitives form the atomic input spaces for subsequent test generation.
\textsc{GetComponents} tracks a \emph{path}: a dot-separated string (\eg,
"forecast.hourly.0") that records the location of each primitive from the
root of the structure. 

\begin{algorithm}[t]
\caption{Compute the set of primitive components corresponding to type \texttt{t}; the Path $p$ 
tracks location from root, \eg, "forecast.temp".}
\label{alg:getcomponents}
\begin{algorithmic}
    \myalgofonts
    \Function{getComponents}{$t$: Type, $p$: Path} 
    \If{$t$ is Primitive}
        \State $\{ \text{Component}(t, p) \}$
    \ElsIf{$t$ is model}
        \State $\bigcup_{\text{prop} \in \text{t.properties}}$ \Call{getComponents}{prop.type, p + . + prop.name}
    \ElsIf{$t$ is U[]} 
        \State $\{$Component(Nat, p + .size)$\}$ $\cup$ $\bigcup_{i}$ \Call{getComponents}{U, p + .i}
    \Else \Comment{Otherwise it is an abstract type$\ddagger$}
        \State $\bigcup_{\text{tsub} \in \text{t.subtypes}}$ \Call{getComponents}{tsub, p + @ + tsub.name} 
    \EndIf
\EndFunction
\end{algorithmic}
\end{algorithm}

The code in \Cref{alg:getcomponents} is a recursive dispatch on the types. The base-case is for primitive types, which immediately returns a component for the
type with the path taken from the root of the structure to the primitive. For composite \cf{model} types (objects with named properties), the function recursively calls itself 
on each property, appending the property name to the path:  For example, starting with path "forecast", recursing on the property "temp" yields the path "forecast.temp". To handle array types, \projName creates a synthetic component for the length of the list, restricted to $0$--\cf{Max}, and then recursively calls 
\Call{getComponents}{U, p + .i} for each index $i$ (\Cref{sec:decomp:collect}). For abstract types ($\ddagger$), \projName  
recursively calls each possible subtype and extends the path with a synthetic component to track the subtype choice (\Cref{sec:decompsubtypes}).

\subsection{Decomposing Collections}
\label{sec:decomp:collect}

Collection types, such as arrays, present challenges in the form of their potentially unbounded sizes and the dependencies they introduce between choice 
of sizes and element values. 

In our sample code (\Cref{fig:api-spec-example}), we have an \cf{hourly} field that is an array of integers representing the percentage chance of precipitation 
for each hour of the day. To decompose this type, we first need to consider the size of the array. In theory, we could treat this value uniformly like the other 
primitive types. In practice, however, the larger we allow this array to be, the more input components it introduces, which can quickly trigger a combinatorial explosion 
in the number of tests generated. To limit this explosion, we 
leverage the sparsity of effects principle (\Cref{sec:soep}) and 
limit the choices for the sizes of collections to be in the range of $0$ to $3$ elements.

In addition to bounding the size of the collection, we also consider the dependency between the size of the collection and the valid index values 
for the components it includes. Specifically, we use two consistency constraints, a bounds check for collections (here) and a type check for subtyping (\Cref{sec:decompsubtypes}), and automatically 
tag the input components with them as needed during the decomposition process.
An example of this is shown in \Cref{tab:temp-range-values}, which demonstrates how constraints are used to prevent infeasible pairings 
such as generating a test case where a array is empty (\cf{v.hourly = []}) but also attempting to assign a value to the first element 
(\cf{v.hourly[0] = 50}).

\begin{table}[t] 
\caption{Example values for the \cf{v.hourly} array input and elements (with constraints).}
\label{tab:temp-range-values}
\centering
\ttfamily
\footnotesize
\begin{tabular}{lc}
\toprule
\textbf{Parameter} & \textbf{Values} \\
\midrule
\cf{v.hourly.size} & $\lbrace 0, 1, 2, 3 \rbrace$ \\
\midrule
\cf{v.hourly.0.value} & $\lbrace 0, 10, 0 \rbrace$ $\land$ $\cf{v.hourly.size} > 0$ \\
\cf{v.hourly.1.value} & $\lbrace 10, 20 \rbrace$ $\land$ $\cf{v.hourly.size} > 1$ \\
\cf{v.hourly.2.value} & $\lbrace 80 \rbrace$ $\land$ $\cf{v.hourly.size} > 2$ \\
\bottomrule
\end{tabular}
\vspace{3mm}
\vspace{-5mm}
\end{table}

The generated components and possible values for test generation are shown in \Cref{tab:temp-range-values}. The first row shows the component representing the size 
of the array, which is always fixed to have the values $\{0, 1, 2, 3\}$. The next three rows show possible values for each element in the array, along with the 
automatically generated constraints on when they can be selected. In this case, if we select the component \cf{v.hourly.1.value}, then we have the 
constraint that we must also select \cf{v.hourly.size} as greater than $1$.

\begin{table}[t] 
    \caption{Example values for the \cf{v.forecastInfo} (\cf{v.fI}) entity input and elements (with constraints).}
\label{tab:forecast-info-values}
\centering
\ttfamily
\resizebox{\columnwidth}{!}{
\begin{tabular}{lc}
\toprule
\textbf{Parameter} & \textbf{Values} \\
\midrule
\cf{v.fI.type} & $\lbrace \text{Sunny}, \text{Cloudy}, \text{Precip} \rbrace$ \\
\cf{v.fI.Precip.stormWatch} & $\lbrace \cf{t}, \cf{f} \rbrace$ $\land$ $\cf{v.fI.type} = \text{Precip}$ \\
\bottomrule
\end{tabular}
}
\vspace{3mm}
\vspace{-7mm}
\end{table}

\subsection{Decomposing Subtypes and Inductive Types}
\label{sec:decompsubtypes}

The decomposition of components that may represent multiple subtypes or inductive types requires additional care. In the case of 
subtypes, we can decompose each possible subtype independently. Then for the path, we add a synthetic component that represents the 
subtype selection. As with the collections, this also requires adding constraints on when each subtype can be selected.

Inductive types, such as trees, are handled by imposing a fixed depth bound. This ensures that the test cases remain compact and, as with collections, 
has a small impact on test effectiveness in practice 
under the sparsity of effects principle.

The component decomposition and possible value selection for the abstract type \cf{ForecastInfo} (from \Cref{fig:api-spec-example}) is shown 
in \Cref{tab:forecast-info-values}. As shown, the \cf{v.forecastInfo.type} component is a synthetic component that represents the subtype selection. 
For subtypes that do not have any additional fields (e.g., \cf{Sunny} and \cf{Cloudy}), we do not generate any additional components.
For the \cf{Precip} subtype, we generate an additional component for the \cf{stormWatch} field and add a constraint to ensure 
this component is only selected when the \cf{v.forecastInfo.type} is \cf{Precip}.

\section{LLM-Guided Partitioning and Generation}
\label{sec:llm-partitions}

Once \projName has recursively decomposed an ADT in a test input into its set of primitive components, the next step is to partition those primitives and select representative test values. \projName realizes this step via \textsc{GenerateValues} (Algorithm~\ref{alg:gen-tests}). Traditionally, black-box testing relies on domain experts to manually partition the input space~\cite{combtesting}. While LLMs can automate this, existing frameworks generate tests in a two phases: first finding a sequence of calls to build the requisite server-side state, and then generating a complex client-side payload for that state in a single monolithic request --- a process known to be brittle and expensive.

In contrast, our core technical novelty lies in a fundamental paradigm shift: \projName is the first framework to isolate the LLM's scope to individual primitive components, using the model strictly as a semantic equivalence class generator integrated within a symbolic combinatorial engine. By avoiding monolithic payload generation, this targeted application yields higher-quality test suites, stable coverage, low inference costs, and robustness to model changes (\Cref{sec:eval}). We realize this novel approach by contextualizing these isolated primitives at two distinct layers: local semantic context (\Cref{sec:local-context}) and global structural context (\Cref{sec:global-context}).

\subsection{Local Context}
\label{sec:local-context}

In the absence of context, an LLM cannot do much better than randomly sampling the domain for a given type. At first this may appear to be a problem, 
as the primitive components simply contain the (primitive) type of the component and a path. However, a simple traversal of the path provides a wealth of 
information about what a particular type (\cf{int32}, \cf{string}, $\ldots$) is intended to represent.

To enable an LLM to generate more meaningful partitions and values, we start by providing it with the local surrounding context. The \cf{TempRange} running 
example from \Cref{fig:api-spec-example} and the primitive components at \cf{v.low.value} and \cf{v.high.value} demonstrates the forms of local context that 
we leverage to build a local context for the LLM prompt. 

\begin{table}[t]
\caption{Generated values for \cf{v.low.value} and \cf{v.high.value} in \cf{TempRange} using \emph{No} or \emph{Local} context.}
\label{tab:temp-range-llm-example}
\centering
\ttfamily
\footnotesize
\begin{tabular}{lcc}
\toprule
\textbf{Parameter} & \textbf{No Context} & \textbf{Local Context} \\
\midrule
\cf{v.low.value} & $\lbrace -1, 0, 5 \rbrace$ & $\lbrace -10, 0, 32, 70, \ldots \rbrace$ \\
\cf{v.high.value} & $\lbrace -1, 0, 3 \rbrace$ & $\lbrace -10, 0, 32, 70, \ldots \rbrace$ \\
\bottomrule
\end{tabular}
\vspace{3mm}
\vspace{-7mm}
\end{table}

As shown in \Cref{fig:local_llm_context}, the first piece of context we use is that the \cf{int32} value is directly inside a \cf{scalar} alias 
declaration; the template-terms in the prompt are enclosed in \cf{\{\{$\ldots$\}\}}. These declarations provide key semantic information about the intent of the underlying value. 
In this case, that the underlying \cf{int32} is a temperature in Fahrenheit. A LLM can leverage this information to partition the input space for the value and determine which 
inputs represent interesting test cases. In some cases, like the \cf{Percentage}, we also have explicit invariants, in the form or code decorators or 
regular expression constraints for strings\footnote{\lstinline{@pattern("/[0-9]\{5\}(-[0-9]\{4\})?/") scalar 
Zipcode extends string;} is a type alias of a string representing a zipcode} that hold on the value.

\begin{figure}[t]
    \footnotesize
    \centering
    \begin{tcolorbox}[
        colback=white, 
        colframe=black, 
        boxrule=1pt, 
        arc=4mm, 
        title=Local LLM Prompt Context, 
        fonttitle=\bfseries 
    ]
    Given a property, named \{\{low\}\} in a type named \{\{TempRange\}\}, with data type \{\{int32\}\}, 
    generate a JSON array containing ONLY test values strictly matching the specified data type and format, covering all possible interesting input value equivalence classes,
    boundary conditions, and unusual or unexpected values for testing.
    \end{tcolorbox}
    \vspace{-2mm}
	\caption{LLM local context prompt for \cf{v.temp.low.value} input generation. The \cf{\{\{$\ldots$\}\}} text are where template positions in the 
    prompt have been instantiated with values from the running example. This is concatenated with additional components are appended to produce the final prompt.}
    \label{fig:local_llm_context}
    \vspace{-4mm}
\end{figure}

The second piece of context we use is information on the variable or enclosing type + property (as applicable) that the value is associated with. As with the type alias 
the name of the variable or type/property name provide rich context about the intent and likely usage of the value. In our example of the \cf{TempRange} entity, the 
type indicates that we expect a range in values while the specific property name (\cf{low} or \cf{high}) provides additional context about the intended usage of the 
value and possible biases --- \eg towards low or high ends of the expected ranges.

\Cref{tab:temp-range-llm-example} shows the result of querying our LLM with and without the local context information. As can be seen in the 
no-context column the results without additional context are simply a selection of values that cover three major partitions of a generic integer, negative, zero, and positive. 
These are clearly not sufficient for robust testing and will probably miss interesting behaviors, \eg $32$ degree freezing point transition. 
However, with the additional information in the local context prompt, the LLM is able to generate values that are much more representative of the 
expected temperature input space as shown in column three.

\subsection{Global Context}
\label{sec:global-context}

Despite the improved generation of values with local context, and the much better coverage of the input spaces, these values can miss critical context captured in 
the larger API structure. For example, in \Cref{tab:temp-range-llm-example}, the low and high values both end up with the same distribution when the, likely, intent 
is that the low value is less than the high value. In this case, the result is simply inefficiency, \eg many redundant or invalid test cases are generated, 
but it can also lead to poor coverage as well. Imagine that instead of a weather forecast, we are working with a temperature control system for an engine (or 
liquefied gasses). In this scenario, picking temperatures corresponding to ambient values will completely miss much larger (or smaller) representative values.

\begin{figure}[t]
    \footnotesize
    \centering
    \begin{tcolorbox}[
        colback=white, 
        colframe=black, 
        boxrule=1pt, 
        arc=4mm, 
        title=Global LLM Prompt Context, 
        fonttitle=\bfseries 
    ]
    The API that this value is being generated for has the following signature:\\
    \{\{op recommendedActivities(v: Forecast): string[]\}\}\\
    
    The value is on the path:\\
    \{\{v.low.value\}\}\\
    
    The traversed type definitions are: \{\{\\
    model Forecast \{ temp: TempRange; ... \}\\
    model TempRange \{ low: Fahrenheit; high: Fahrenheit; \}\\
    scalar Fahrenheit extends int32;\\
    \}\}
    \end{tcolorbox}
    \vspace{-2mm}
    \caption{LLM global context prompt for \cf{v.temp.low.value} input generation. The \cf{\{\{$\ldots$\}\}} text are where template positions in the 
    prompt have been instantiated with values from the running example. This is concatenated with additional components are appended to produce the final prompt.}
    \label{fig:global_llm_context}
    \vspace{-5mm}
\end{figure}

\begin{table}[t]
\caption{Generated values for \cf{v.low.value} and \cf{v.high.value} in the \cf{TempRange} using \emph{Global} context.}
\label{tab:temp-range-llm-example-global}
\centering
\ttfamily
\footnotesize
\begin{tabular}{lc}
\toprule
\textbf{Parameter} & \textbf{Global Context} \\
\midrule
\cf{v.low.value} & $\lbrace -10, 0, 32, 60, \ldots \rbrace$ \\
\cf{v.high.value} & $\lbrace 0, 32, 70, 95, 110, \ldots \rbrace$ \\
\bottomrule
\end{tabular}
\vspace{3mm}
\vspace{-7mm}
\end{table}

To counteract this, we generate global context that provides the LLM with information about the overall environment in which a primitive component operates. 
This global context includes the top-level signature of the API under test and the full path of types and fields to the primitive component 
(including doc comments).

This context that is generated and appended to the general LLM prompt is shown in \Cref{fig:global_llm_context}. The 
global context provides the LLM with information about the semantics and intended overall roles of the values that we want to generate. Specifically, 
that the values are (low) temperatures in Fahrenheit, as part of a \cf{Forecast} in an API for recommending activities. 

\Cref{tab:temp-range-llm-example-global} shows the result of querying our LLM with the global \emph{and} local context information. 
The LLM leverages the additional information in the global context prompt to infer that the temperatures are ambient weather 
values as well as capture the intent of the low and high fields, enabling it to bias these values appropriately.

The final part of the prompt specifies the agentic instructions. These are designed to frame problem we are trying to address clearly for the LLM and to 
provide direct instructions on how to proceed with the information that is provided. This prompt is constructed around standard design rules and structure 
for LLM prompts across models and versions~\cite{anthropicguide}.

\section{Sparsity-Guided ADT Generation}
\label{sec:soep}

Given the set of representative values for the primitive components of an ADT, computed in \Cref{alg:getcomponents} and \Cref{sec:llm-partitions}, the next step is to 
recombine these values into test cases. The direct approach would be to select values for each component independently at random and combine them into a test case. However, 
like all randomized selection approaches, this technique suffers from the \emph{coupon collector's problem} and as the number of generated tests grows the likelyhood of 
producing new and useful tests decreases. Direct application of an LLM to perform the combination is also limited in effectiveness as context window sizes and costs require 
either super-linearly increasing token costs\footnote{Our engineering prototype experiment indicated an order-of-magnitude increase in cost as well as lower quality test-suites 
\vs the technique described in \Cref{alg:gen-comb-test}.} and attention degradation or memoryless (and again coupon-collector-like) sampling of values for each component. 

To address this problem, \projName uses the Sparsity-of-Effects Principle (SoEP)~\cite{kuhn2010practical} to guide the recombination of values into test cases. SoEP is a 
well-established principle in testing that posits that most software failures are triggered by interactions between a small number of parameters. This principle 
has appeared in many forms in the literature, including the small-model (or small-scope) hypothesis~\cite{smallscope} in model-checking literature, the software practitioners 
maxim's that ``all bugs are shallow'' or that bug reports should include a ``small reproduction'', and in the combinatorial testing literature is often called the 
small interaction rule~\cite{kuhn2010practical}. By leveraging this principle, \projName focuses on generating test cases that cover interactions between a small number of 
components, and LLM generated values for these components, which are more likely to reveal faults.

Given a set of primitive components, each with a set of candidate values, the \textsc{GenKWay} function (\Cref{alg:gen-comb-test}) generates a set 
of covering tests. These are built from \emph{k-way} combinations of generated values ($\dagger$ in \Cref{alg:gen-tests}) over the selected components 
and are augmented with a random selection of values for the remaining components. 
The outer-loop ensures that each possible combination of representative values for the selected components are explored. The inner-loop constructs a 
complete test value by iterating over all components. If the component is one of the selected components, then the value from the current combination is used. 
Otherwise, a value is selected uniformly at random from the candidate values.

\begin{algorithm}[t]
    \caption{\textsc{GenKWay}: Test construction via full $k$-way primitive component sampling ($c_1$, $\ldots$, $c_k$) with loose sampling over remaining components (allc).}
\label{alg:gen-comb-test}
\begin{algorithmic}
    \myalgofonts
    \Function{GenKWay}{$c_1$, $\ldots$, $c_k$: \lstinline[basicstyle=\tiny\ttfamily]+Set<Component>+, allc: \lstinline[basicstyle=\tiny\ttfamily]+Set<Component>+} 
    \State ktests $\gets$ $\emptyset$
	\ForAll{$v \in c_1.\mathit{vals} \times \ldots \times c_k.\mathit{vals}$} \Comment{All \emph{partitioned} combos}
        \State test $\gets$ $\emptyset$
        \ForAll{c$_x$ $\in$ allc}
            \If{$\exists i \in [0, k]$ such that $c_x = c_i$} \Comment{Combinatorial}
                \State test $\gets$ test $\cup$ \{ $c_x.path = \mathbf{v}[i]$ \} \Comment{$i$-th value from tuple}
            \Else 
                \State vselect $\gets$ \Call{sample}{c$_x$.values} \Comment{Uniform selection}
                \State test $\gets$ test $\cup$ \{c$_x$.path $=$ vselect\}
            \EndIf
        \EndFor
        \State ktests $\gets$ ktests $\cup$ test
    \EndFor
    \State \Return ktests
\EndFunction
\end{algorithmic}
\end{algorithm}

Applying this approach to the \cf{TempRange} example from \Cref{fig:api-spec-example} and the generated values from \Cref{tab:temp-range-values}, we can generate a set of $1$-way 
combinatorial tests that cover all possible values for the \cf{low} \emph{or} \cf{high} fields independently (totaling $6$ tests) as shown in \Cref{tab:temp-range-one-way}. 
Alternatively we can generate a set of $2$-way combinatorial tests that cover all possible value combinations for the \cf{low} \emph{and} \cf{high} fields (totaling $9$ tests) as 
shown in \Cref{tab:temp-range-test-cases}. As we increase the value of $k$, we can generate tests that cover more complex interactions between the components, but this also leads 
to an exponential increase in the number of tests generated -- for example the test case $\lbrace 42, 80 \rbrace$ is generated in the $2$-way case but not in the $1$-way case. 

\begin{table}[t]
\caption{Example test cases for the \cf{TempRange} model (record) by 1-way sampling
$\lbrace -10, 0, 42 \rbrace$ and 
$\lbrace 32, 60, 80 \rbrace$.
}
\label{tab:temp-range-one-way}
\centering
\ttfamily
\footnotesize
\begin{tabular}{ccc}
\toprule
\textbf{v.low.value} & \textbf{v.high.value} & \textbf{TempRange} \\
\midrule
\bf{-10} & 32 & $\lbrace -10, 32 \rbrace$ \\
\bf{0} & 60 & $\lbrace 0, 60 \rbrace$ \\
\bf{42} & 32 & $\lbrace 42, 32 \rbrace$ \\
42 & \bf{32} & $\lbrace 42, 32 \rbrace$ \\
-10 & \bf{60} & $\lbrace -10, 60 \rbrace$ \\
-10 & \bf{80} & $\lbrace -10, 80 \rbrace$ \\
\bottomrule
\end{tabular}
\end{table}

Empirically, $2$-way combinatorial testing is generally considered to be capable of revealing at least $60\%$ of faults in a system while $3$-way and $4$-way combinatorial 
testing can reveal on the order of $80\%$ and $95\%$ respectively~\cite{combtesting,nie2011survey} at rapidly increasing test costs. This technique has traditionally been 
applied to systems with a fixed set of primitive typed input parameters and base values, often these values are selected by a domain expert, that 
are designed to cover all interesting regions for each parameter. However, due to the need for domain expertise in constructing the test values, and the difficulty of defining 
partitions for complex data types, combinatorial testing has primarily been used in safety-critical or high-assurance applications with simple (flat) inputs.
As we show in \Cref{sec:eval}, and specifically the ability to uncover faults that are not revealed by prior test generation approaches using the $2$-way sampling\footnote{Preliminary 
experiments with $k > 2$ sampling resulted in excessively large test suites even on smaller benchmarks. As as result we leave the problem of effectively applying higher $k$-way or 
dynamic-way sampling to future work. We also note the extensive literature on covering array algorithms~\cite{coveringarraysurvey} from the combinatorial testing community that 
could be applied here as well.}, this technique generalizes effectively to the structured input generation problem for complex ADTs and our novel approach as well.

\section{Mock-Aware LLM Value Generation}
\label{sec:mockgen}

Generating tests for code that interacts with other data-sources or systems is a vexing problem. In some cases, developers simply run tests against live systems, 
risking non-deterministic/flaky tests and possible data corruption. In other cases, developers use mocks~\cite{declmock,mockingstudy} or stubs to simulate interactions 
with external systems. These mocks ensure deterministic behavior and, particularly in the case of external networks calls, can substantially speed up test execution. 
However, mocks present a substantial challenge for test generation. Consider a mock database deployed to support testing an 
application that manages a set of customer records (\Cref{tab:person-mock}).

If we are testing an API that interacts with this database, then the effectiveness of the generated test suite will depend 
on how well the generated tests interact with this mock data~\cite{mockingstudy}. The technique used for LLM-based value 
generation in \Cref{sec:llm-partitions} struggles with this task, as the prompts only include the structure of the data and types 
of values --- \ie, and \cf{id} is a \cf{UUID}. While the LLM can generate many valid UUIDs, it is very unlikely to generate the specific 
UUID present in the mock database, greatly reducing test coverage.


\begin{table}[t]
\caption{Sample data entries in Mock database for testing.}
\label{tab:person-mock}
\centering
\ttfamily
\resizebox{\columnwidth}{!}{%
\begin{tabular}{llll}
\toprule
\textbf{id} & \textbf{name} & \textbf{age} & \textbf{createdOn}\\
\midrule
696f0b92-7477-4ced-a7ef-9e63038b9fc0 & Steve & 27 & 2024-01-31 \\
bb4d6e69-5be2-488c-aef0-fc0627d40cf4 & Alice & 25 & 2021-09-16 \\
55a62005-0c72-4dd2-a9a6-239d9008c828 & Bob & 22 & 2025-02-26 \\
37f8a128-4a0b-423c-8be3-eb13bae56554 & John & 30 & 2025-06-16 \\
\bottomrule
\end{tabular}%
}
\vspace{3mm}
\vspace{-5mm}
\end{table}

To address this problem \projName applies two techniques. The first is when the developer has already implemented a mock datasource and we 
merely need to integrate this data into the test generation process. The second is the case where the developer has not yet implemented a mock datasource
and we want to automatically provision this data source.

In the first case \projName performs a walk of the codebase looking for any mock data sources, such as databases, files, or mock 
libraries, and extract the relevant textual data. During the value generation process (\Cref{sec:llm-partitions}), we inject a sampling (based on 
data size and model context windows limits) this data directly into the LLM prompt.

In the case where the analysis of the codebase does not include an implemented mock, but does contain database-description tables, XML schema, or other 
data formats, \projName uses the LLM to generate mock data based on the structure of the data and the types of values via decomposition and $1$-way sampling. 
This generated mock data is then injected into the prompt in the same way as described above. As shown in \Cref{sec:eval}, this approach is effective at 
generating test values that interact with the mock data, improving test coverage by up to $20\%$ in our feature ablation study.

\section{Evaluation}
\label{sec:eval}
This section presents the results of the empirical studies conducted to assess \projName. Our evaluation aims to
address the following research questions:
\begin{description}
\item[RQ1:] How does \projName compare with state-of-theart REST API testing tools in terms of code coverage and
operation coverage achieved?
\item[RQ2:] In terms of error detection, how does \projName perform in triggering distinct 500 (Internal Server Error) responses 
compared to state-of-the-art REST API testing tools?
\item[RQ3:] How do the sub-components of \projName (decomposition, LLM value generation, and mock-integration) contribute to 
its overall performance?
\item[RQ4:] What is the cost, both CPU time and LLM inference costs, of test generation and how sensitive is the technique 
to the underlying LLM used?
\end{description}

\subsection{Methodology}
The evaluation is based on a set of publicly available~\cite{arat-github} RESTful benchmarks introduced by Kim \etal~\cite{arat-rl} and 
subsequently extended in~\cite{autoresttest}. These benchmarks have been used in multiple recent studies~\cite{autoresttest,arat-rl,emtest,llamaresttest} on SotA RESTful API 
testing and provide a standard baseline for comparison. To support reproducibility and 
enable analysis of discovered errors, we exclude three services (FDIC, Spotify, and Ohsome API) which are closed-source. We compare \projName against 
existing SotA testing tools ARAT-RL~\cite{arat-rl}, EvoMaster~\cite{evomaster}, and AutoRestTest~\cite{autoresttest}, as well as the 
industry standard RESTler~\cite{restler}.

\projName focuses solely on generating tests for RESTful endpoints in isolation, \eg each test is a single 
call to a single endpoint. Additionally, the test suite is generated entirely offline --- 
as opposed to prior work which generates tests online. Thus, for all experiments we provide \projName with the API specification 
and any mock data sources used by the service under test. \projName then generates a test suite for each service, writing all the 
test files to disk. The service under test is then started in a clean state and the generated tests are executed sequentially,
for the GET, POST, PATCH, PUT, and DELETE endpoints. In cases where a service requires authentication this API is run first and, if \projName cannot 
successfully validate, then the testing is halted.

\addlater{Internally, \projName translates each benchmark's TypeSpec (or OpenAPI) specification into a \bosque~\cite{bsqon,bosque} internal representation for test generation.}
We conducted our experiments on a workstation equipped with an AMD 16 core Ryzen™ 9 9950X processor, 128 GB of RAM, and coverage measurements 
are taken with the Jacoco framework~\cite{jacoco}. To avoid spurious coverage results, we configure Jacoco to exclude all test and test-support files. 
The default LLM model used for \projName test generation is deepseek-chat (DeepSeek-V3.1) in non-thinking mode. 

All tables report the median coverage values or average number of distinct 500 errors found per test generation run -- over $10$ runs. 
The maximum variance in coverage results across \projName runs is $3\%$ in line/branch coverage and the same set of 500 errors was 
detected on every run.

\begin{table}
\caption{RESTful service benchmarks statistics -- LOC and number of API endpoints provided by each service.}
\label{tab:benchmarks}
\centering
\footnotesize
\setlength{\tabcolsep}{6pt}
\begin{tabular}{lrr}
\toprule
REST Service      & Code Lines & Endpoints \\
\midrule
Genome Nexus      & 22.1k      & 23 \\
Market            & 7.9k       & 13 \\
Person Controller & 0.6k       & 12 \\
User Management   & 2.8k       & 22 \\
Features Service  & 1.6k       & 18 \\
YouTube-Mock      & 2.4k       & 1\\
Project Tracking System   & 3.7k      & 67 \\
Language Tool            & 113.2k       & 2 \\
Rest Countries & 1.6k       & 22 \\
\bottomrule
\end{tabular}
\vspace{3mm}
\vspace{-7mm}
\end{table}

\subsection*{RQ1: Code Coverage Evaluation}
\label{sec:topline}

\begin{figure*}[t]
    \centering
    \includegraphics[
        width=0.90\textwidth,
        keepaspectratio,
    ]{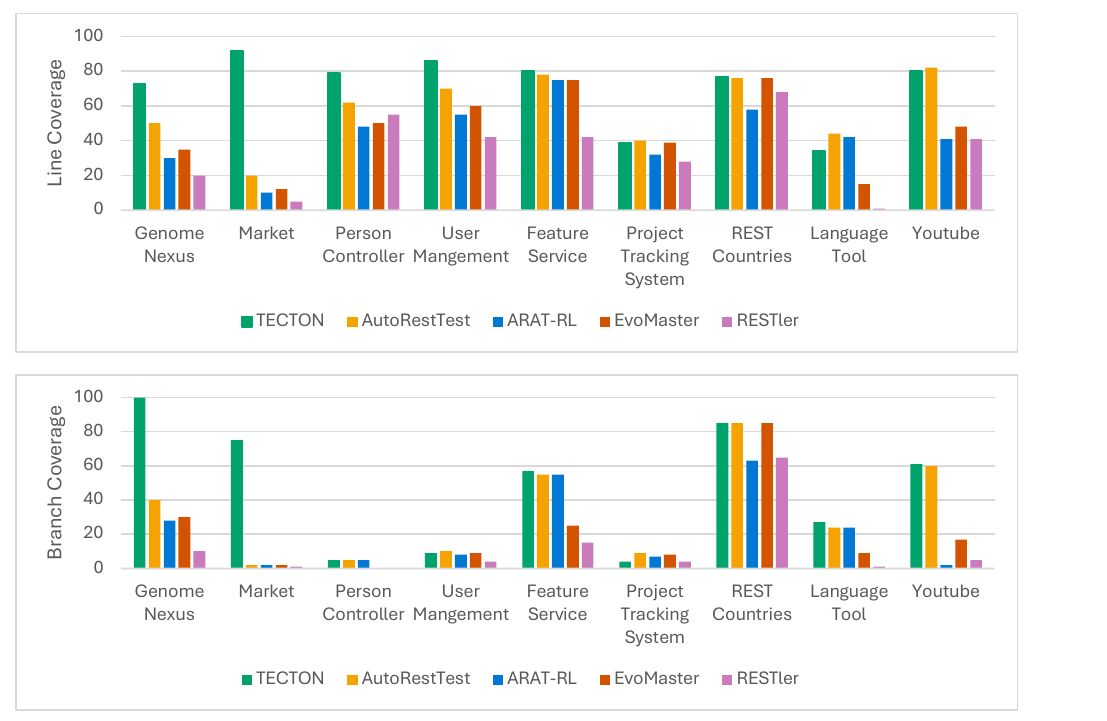}
    \caption{Coverage evaluation, line and branch, on all benchmarks using \projName along with, AutoRestTest, ARAT-RL, EvoMaster, and RESTler. Top figure shows line coverage, with \projName achieving best results in $6/9$ cases. Bottom figure shows branch coverage, with \projName achieving best results in $5/9$ cases and near parity in the remaining cases.}
    \label{fig:TGenEval}
\end{figure*}

\begin{table}[t]
\caption{Each column reports number of distinct service failures triggered (HTTP 500 response codes) for each technique across all services. \projName triggers more than 
twice as many service failures as the other tools, exposing previously undetected failures in 5 out of 9 services, and missing only one 500 error for each of the Language Tool and YouTube applications.}
\label{tab:service_failures}
\centering
\resizebox{\columnwidth}{!}{
\begin{tabular}{lrrrrr}
\toprule
\textbf{REST Service} & \multicolumn{1}{c}{\textbf{\projName}}      & \multicolumn{1}{c}{\textbf{AutoRestTest}} & \multicolumn{1}{c}{\textbf{ARAT-RL}}    & \multicolumn{1}{c}{\textbf{EvoMaster}}  & \multicolumn{1}{c}{\textbf{RESTler}} \\
\midrule
Features Service        &9 & 1  & 1  & 1  & 1 \\
Language Tool           &0 & 1  & 1  & 1  & 0 \\
REST Countries          &1 & 1  & 1  & 1  & 1 \\
Genome Nexus            &7 & 1  & 1  & 0  & 0 \\
Person Controller       &9 & 8  & 8  & 8  & 3 \\
User Management         &1 & 1  & 1  & 1  & 1 \\
Market                  &5 & 1  & 1  & 1  & 1 \\
YouTube                 &0 & 1  & 1  & 1  & 1 \\
Project Tracking System &2 & 1  & 1  & 1  & 1 \\
\midrule
\textbf{Total}          & \textbf{34} & \textbf{16} & \textbf{16} & \textbf{15} & \textbf{9} \\
\bottomrule
\end{tabular}
}
\vspace{5mm}
\vspace{-7mm}
\end{table}

As shown in \Cref{fig:TGenEval}, \projName consistently outperforms previous techniques in code coverage metrics. In particular, \projName achieves high 
line coverage, over 70\%, across all benchmarks except \emph{Project Tracking System} and \emph{Language Tool}, whose coverage are low for all systems, 
and substantially outperforms the previous state-of-the-art by $10\%$ or more for $4$ applications. 
\projName also outperforms previous techniques in branch coverage in $5$ benchmarks and is near parity in the remaining $4$. 
\addlater{The topline results for branch coverage in \emph{Person Controller} and \emph{User Management} are notably low for all techniques. We explore these 
results in more detail in \Cref{sec:covlimit}.}
These results show that \projName can effectively explore REST services, achieving superior code
coverage when compared to existing tools, and demonstrates its potential for REST API testing

\subsection*{RQ2: Fault-Detection Capability}
The results in \Cref{tab:service_failures} show the number of distinct service failures (HTTP 500 response codes) triggered by each tool across all benchmarks. 
In total, \projName triggers $34$ (manually verified) distinct service failures, more than twice as many as the next best tools, AutoRestTest and ARAT-RL, which 
each trigger $16$ failures. This demonstrates \projName's effectiveness in generating high-quality test cases that can uncover critical issues in RESTful services.

As hypothesized in \Cref{sec:intro}, these results demonstrate that by leveraging the sparsity-of-effects principle~\cite{PNUELI2002279,smallscope},
\projName is able to generate tests that are both diverse and that effectively uncover faults. The higher number of distinct service failures triggered 
when compared to other tools shows that \projName is able to explore the input space more effectively, leading to the discovery of previously undetected 
failures in $5$ out of $9$ services.

\subsection*{RQ3: Ablation Study of \projName Components}
\label{sec:ablation}
\begin{figure}[t]
    \centering
  	\includegraphics[width=.90\linewidth]{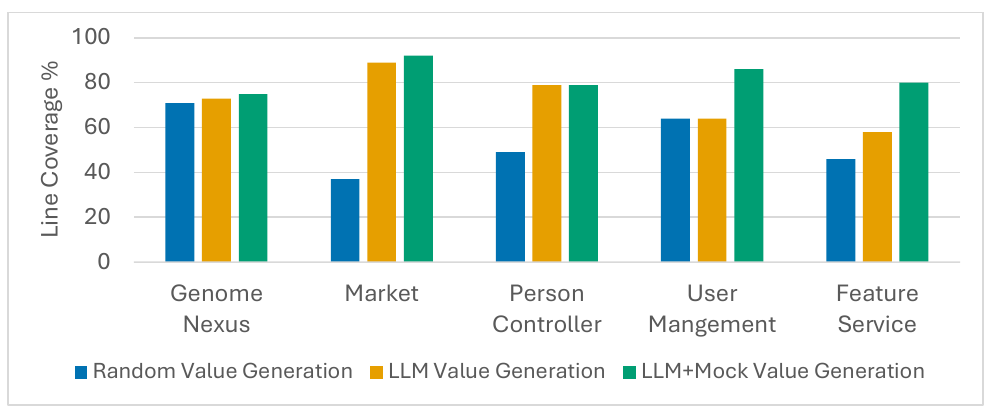}
	\caption{Ablation study to quantify contributions of the \projName components (on line coverage). Configurations include $k$-way from 
    \Cref{sec:soep} with random value generation for the primitive partitions (\emph{Random Value Generation}), sampling test generation 
    with LLM-based value generation from \Cref{sec:llm-partitions} (\emph{LLM Value Generation}), and the full \projName approach including mock-awareness 
    (\emph{LLM+Mock Value Generation}).}
    \label{fig:TGenEvalAblation}
    \vspace{-5mm}
\end{figure}

\Cref{fig:TGenEvalAblation} evaluates the impact of the individual components of the \projName test generation methodology.
The first column, \emph{Random Value Generation}, uses $2$-way sampling from \Cref{sec:soep} 
and random value generation to populate primitive partitions. The second column, \emph{LLM Value Generation}, uses the sampling test 
generation approach from \Cref{sec:decompose} and the LLM-based value generation from \Cref{sec:llm-partitions} to populate primitive partitions. 
The third column, \emph{LLM+Mock Value Generation}, is the full \projName approach which includes mock-aware test generation from \Cref{sec:mockgen}.

Each component contributes meaningfully to the overall performance of \projName, the standalone combinatorial test generation 
with random value generation (\emph{Random Value Generation}) provides a solid baseline, ranging from $30\%$-$70\%$ line coverage, and even out-performing 
previous state-of-the-art on the \emph{Genome Nexus} benchmark. Introducing LLM-generation provides a universal and in $3/5$ tests a substantial boost in coverage.
Finally, the introduction of mock-aware generation provides a further boost in coverage 
in critical cases. In the cases of the \emph{User Management} and \emph{Feature Service} benchmarks, the addition of mock-context information results in an 
$20\%$+ increase in coverage. These programs use specific identifiers from a prepopulated testing database
to exercise various code paths and, without access to this explicit mock-data context, the LLM can generate a valid identifier but not one that 
is actually present in the database.

\subsection*{RQ4: Test Generation Cost and Sensitivity to LLM Choice}
\label{sec:costs}
For runs with the default LLM model, deepseek-chat, the timeout for test generation was set to $1$ hour as 
in previous work. However, the largest test generation time observed was 35 minutes. LLM inference costs never 
exceed USD $\$0.10$ ($700$k tokens) for any benchmark. These results show that by targeting the LLM to the 
specific aspect of generation where it's abilities are most effective, \projName is able to generate 
high-quality test suites in a reasonable amount of time and at a low cost (or even completely locally).

To evaluate \projName's sensitivity to the
choice of LLM, we re-ran our coverage experiments on the \emph{Market} application, which has a variety
of input features/types and elicited a large range of reported coverages in prior work. 
We experimented with four alternatives to cover models from different providers that are 
1) a recent/advanced foundation model with GPT-5, 2) models of similar generations to deepseek-chat but 
different families with GPT-4o and Gemini-2.0-flash, and 3) a mid-size open-source model with GPT-OSS 20B.

\begin{figure}[t]
    \centering
    \includegraphics[width=.90\linewidth]{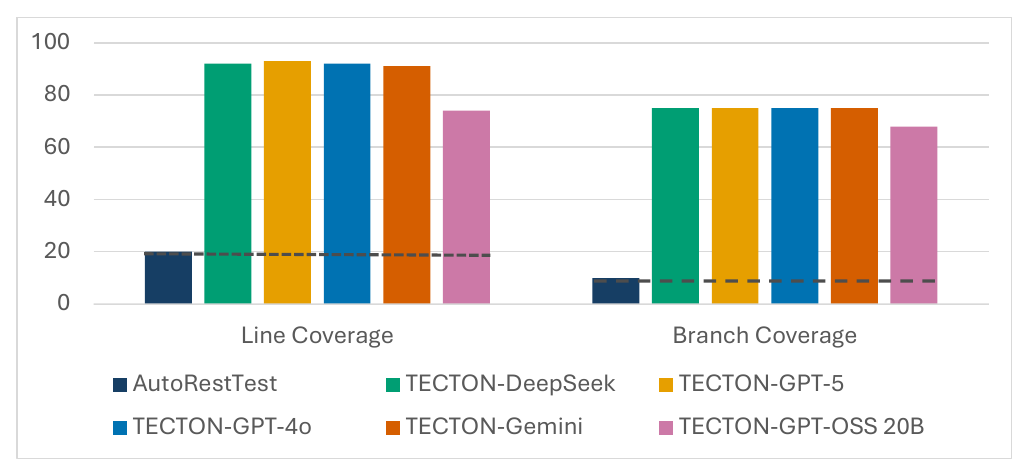}
	\caption{Comparison of \projName using foundation (GPT-5, GPT-4o, Gemini-2.0-flash) and mid-scale open-source (GPT-OSS 20B) 
    LLM's for primitive partition value generation (same prompts). These results, on the \emph{Market} benchmark, demonstrate \projName’s robustness to model 
    choice. All models, including GPT-OSS 20B, surpass previous state-of-the-art methods.}
    \label{fig:models_comparison}
    \vspace{-5mm}
\end{figure}

\Cref{fig:models_comparison} shows the results for the four alternative models, GPT-5, GPT-4o, Gemini-2.0-flash, 
and GPT-OSS 20B, along with the baseline results from deepseek-chat and the previous state-of-the-art results AutoRestTest~\cite{autoresttest}. As can be seen,
all of the foundation models, GPT-5, GPT-4o, Gemini-2.0-flash and DeepSeek, achieve similar coverage results with only minor variations (less than 5\%) 
in coverage metrics. Using the smaller open-source model, GPT-OSS 20B, results in lower coverage, a reduction of $18\%$ line and $7\%$ branch coverage,
but still outperforms the previous state-of-the-art results by $54\%$ and $66\%$ respectively.

\addlater{
\subsection{Effective Coverage Limit}
\label{sec:covlimit}

The final evaluation is a manual analysis of the coverage results in the \emph{User Management} and \emph{Person Controller} benchmarks to understand 
the causes for the low branch coverage and how close the generated tests are to optimal coverage \vs a hand-crafted test suite. 
As seen in \Cref{fig:TGenEval}, all the tools produce low coverage \ie branch coverage under $10\%$ on these two benchmarks.
For each uncovered branch reported by Jacoco, we manually investigated whether the branch was truly unreachable code or if there are specific 
API interactions that could result in the execution of the code. Our goal is to understand the delta between the achieved coverage and the 
theoretical upper bound.

Beginning with the \emph{User Management} application, just over $2.8k$ lines of code, we manually analyzed each uncovered 
branch and associated code. This investigation revealed extensive sections of code that were auto-generated via Lombok~\cite{lombok} annotations. 
Lombok generates default method implementations including code for getters/setters, equality, hashcode, and toString methods. As the 
\emph{User Management} benchmark is a small application, we were able to manually inspect the entire codebase, including possible implicit 
calls to hashcode and equals in map/set uses, to check if a branch was truly unreachable. After this manual inspection we identified 
only $8$ uncovered branches that were possibly reachable -- indicating that despite the low coverage reported by Jacoco, the test suite 
generated by \projName was in fact nearly ideal and even a perfect test suite would only marginally increase the coverage amounts.


A similar analysis on the \emph{Person Controller} application revealed that, again, the uncovered code was default implementations of common methods. 
Based on a manual inspection of the code, $600$ lines in total, we confirmed that all of the uncovered branches were in unreachable code -- making 
the effective code coverage $100\%$.

After removing the unreachable code that we had manually identified, we re-computed the \emph{effective code coverage} for the \projName test suites. 
Post removal, \projName's effective branch coverage rose from 1\% to 100\% for \emph{Person Controller} and from 9\% to 93\% for \emph{User Management}. 
}

\section{Related Work}
\label{sec:relwork}

\noindent
{\em LLM Driven Test Generation:}
Large language models (LLMs) have been widely applied to various aspects of test generation~\cite{llmseoverview,llmtestoverview,fuzz4all}. This work has included 
the development of unit-test suites from scratch, augmenting existing test suites, and generating test oracles~\cite{llmunittest,agora,aitestingmeta,coverup,toga,tico}. 
These systems have been deployed, with success, at scale, in industrial settings~\cite{aitestingmeta}.
Recent work, notably LISP~\cite{llmsplit}, has significantly advanced the use
of LLMs for input space partitioning, showcasing the power of generative models
to reason about complex API signatures. However, such approaches often task the
model with synthesizing entire instances of intricate, nested ADTs in a single
step. In contrast, \projName establishes a more sample-efficient paradigm by shifting
from monolithic generation to structural decomposition. By isolating the LLM's
role to the primitive constituent types and applying the sparsity-of-effects
principle, \projName reduces the task of synthesizing complex objects into a series
of tractable, low-dimensional sub-tasks.\\



\noindent
{\em Structured Input Generation:}
RESTful API fuzzing represents a unique point on the spectrum of structured input generation (SIG) techniques. At the most primitive level, these services can 
be treated as taking as input untyped JSON values. Thus, generating inputs for these APIs can be handled using techniques from the 
classic fuzzing literature including random inputs, constant values, mutation, \etc~\cite{afl,evosuite,dart}. However, state-of-the-art 
for API specification is rapidly moving toward the use of much richer type and specification systems including OpenAPI~\cite{openapi}, Smithy~\cite{smithy}, 
TypeSpec~\cite{typespec} and Bosque~\cite{bosque,bsqon}. These systems provide a rich structure for input generation that can be leveraged to increase the effectiveness 
of test generation~\cite{pex,gramfuzz}. 

Work on SIG using context-free grammars (CFG) has been explored in many contexts~\cite{gramfuzz,zest,zellercacm,jsfunfuzz}. More recently the S3 project has applied grammar 
directed input generation (with constraints) to network protocols and services~\cite{zellercacm,zellermining,zellerinv}. In contrast to this paper, this work approaches 
this topic from the viewpoint of using a CFG-based definition of the physical inputs/outputs of the API using a novel DSL (ISLa) instead of from 
the perspective of a logical type system. Industry adoption of TypeSpec/Smithy over grammar-centric (IDL) representations, \eg ASN.1~\cite{asn1} or SOAP~\cite{soap}, 
for API specifications indicate a strong developer preference for the type-centric perspective.\\

\noindent
{\em RESTful Testing:}
Automated testing for REST APIs is an increasingly important area of research and practice~\cite{surveyrest}, as they have become an increasingly prevalent 
software architecture. The seminal RESTler~\cite{restler} work introduced several foundational concepts for testing REST APIs, including inferring producer-consumer 
dependencies. Later work has applied reinforcement learning~\cite{arat-rl,deeprest}, evolutionary algorithms~\cite{evomaster}, combinatorial 
call-sequence parameter exploration~\cite{restct}, and combining static and dynamic analysis~\cite{rbctest} to improve on the state-of-the-art. 

Recent empirical work~\cite{defects4rest} introduced a benchmark of real-world RESTful API defects, along with manual analysis of their causes,
and demonstrates that many cannot be detected by \emph{any} testing system based on OpenAPI specifications alone as these specifications lack the needed 
semantic information to detect the defects. This highlights the need for richer specification systems for APIs, such as TypeSpec~\cite{typespec}, 
which can provide correctness oracles for testing.

\section{Conclusion} 
\label{sec:conclusion}

This paper introduced \projName, a novel methodology that, rather than treating server-side state as 
an implicit side-effect of a sequence, treats it as a first-class and explicitly synthesizable artifact.
Our evaluation demonstrates the effectiveness of \projName in generating diverse and meaningful test cases. 
The experimental evaluation on services used in prior work shows that \projName significantly outperforms 
existing approaches in terms of both code coverage and error detection. 
Further, by using structural ADT decomposition we are able to focus the unique capabilities of LLM's onto 
a narrow, well-defined task, ensuring predictable and cost efficient outcomes.
Thus, this result represents a key advancement in making automated testing practical and effective for 
real-world API development.

\section*{Data Availability}
All code, data, and benchmarks used in this study are publicly archived at \url{https://zenodo.org/records/19139682}. 

\bibliographystyle{IEEEtran}
\bibliography{bibfile}


\end{document}